\documentclass[12pt]{article}
\usepackage{times}
\usepackage{geometry}
\usepackage[utf8]{inputenc}
\usepackage{enumitem,amssymb}
\usepackage{ragged2e}
\usepackage{graphicx}
\usepackage{wrapfig}
\newlist{thematic}{itemize}{8}
\setlist[thematic]{label=$\square$}
\usepackage{pifont}
\newcommand{\cmark}{\ding{51}}%
\newcommand{\done}{\rlap{$\square$}{\raisebox{2pt}{\large\hspace{1pt}\cmark}}%
\hspace{-2.5pt}}

\newcommand{\kepler}{\emph{Kepler}}
\newcommand{\tess}{\emph{TESS}}
\usepackage{natbib}
\usepackage{multicol}
\usepackage{multirow}
\usepackage{verbatim}
\usepackage{adjustbox}
\usepackage{soul}

\usepackage[numbib]{tocbibind}
\settocbibname{References}
\bibpunct{(}{)}{,}{a}{}{,}

\begin{document}
\raggedright
\huge
Astro2020 Science White Paper \linebreak

Tracing the Origins and Evolution of Small Planets using Their Orbital Obliquities
\normalsize

\noindent \textbf{Thematic Areas:} \hspace*{60pt} \done Planetary Systems \hspace*{10pt} $\square$ Star and Planet Formation \hspace*{20pt}\linebreak
$\square$ Formation and Evolution of Compact Objects \hspace*{31pt} $\square$ Cosmology and Fundamental Physics \linebreak
  $\square$  Stars and Stellar Evolution \hspace*{1pt} $\square$ Resolved Stellar Populations and their Environments \hspace*{40pt} \linebreak
  $\square$    Galaxy Evolution   \hspace*{45pt} $\square$             Multi-Messenger Astronomy and Astrophysics \hspace*{65pt} \linebreak
  
\textbf{Principal Author:}

Name:	Marshall C. Johnson
 \linebreak						
Institution:  The Ohio State University
 \linebreak
Email: johnson.7240@osu.edu
 \linebreak
Phone:  614-688-7426
 \linebreak
 
\textbf{Co-authors:} 
George Zhou (CfA), Brett C. Addison (Southern Queensland), David R. Ciardi (NExScI-Caltech/IPAC), Diana Dragomir (MIT/University of New Mexico), Yasuhiro Hasegawa (JPL/Caltech), Eve J. Lee (Caltech/McGill), Songhu Wang (Yale), Lauren Weiss (Hawaii)
  \linebreak

\textbf{Abstract:}

We recommend an intensive effort to survey and understand the obliquity distribution of small close-in extrasolar planets over the coming decade. The orbital obliquities of exoplanets--i.e., the relative orientation between the planetary orbit and the stellar rotation--is a key tracer of how planets form and migrate. While the orbital obliquities of smaller planets are poorly explored today, a new generation of facilities coming online over the next decade will make such observations possible en masse. Transit spectroscopic observations with the extremely large telescopes will enable us to measure the orbital obliquities of planets as small as $\sim2R_{\oplus}$ around a wide variety of stars, opening a window into the orbital properties of the most common types of planets. This effort will directly contribute to understanding the formation and evolution of planetary systems, a key objective of the National Academy of Sciences' Exoplanet Science Strategies report.

\pagebreak
\justifying
\section{Introduction}

Current and ongoing exoplanet surveys  
have resulted in the discovery of thousands of planets and planet candidates. 
The demographics of these exoplanets have yielded clues about their formation and evolution  \citep[e.g.,][]{Mulders15,Fulton17}.
Most of these efforts, however, have focused on the planetary parameters directly observable in the transit surveys: the planetary radii, orbital periods, and stellar host parameters. Further observations of individual planets to characterize them in detail, while more expensive in terms of telescope time, can be extremely valuable. 
These can include measurements of planetary masses \citep[e.g.,][]{Weiss16,PrietoArranz18}, 
atmospheres \citep[e.g.,][]{Sing16,Hoeijmakers18}, or orbital obliquities \citep[e.g.,][]{Winn05,Winn06}. 
Although comparatively expensive and challenging,  
large statistical samples of such measurements can offer unique information on the histories of exoplanet populations as a whole. 

Planets form in protoplanetary disks that are generally assumed to orbit in the same direction as the host star rotates, but planetary orbits (or the disk itself) could be disturbed by a variety of dynamical processes \citep[e.g.,][]{FabryckyTremaine07,Kaib11,Batygin12,R.Barnes15}. These can change the planetary orbital obliquity, that is, the angle between the planetary orbital and stellar rotational angular momentum vectors (which is also referred to as the ``spin-orbit misalignment'' in the literature). 

{\bf Orbital obliquities 
are thus a key tool for learning about the dynamical histories of exoplanets and exoplanetary systems.} They are relatively straightforward to measure, and are impacted by many of the processes that have been proposed to influence planet formation and migration.
Individual obliquity measurements are difficult to interpret; the power of this technique comes through analyzing the {\it distributions} of obliquities with respect to other secondary parameters.
Within the next decade we will have the capability to measure the orbital obliquities of a large population of small planets, and use these to learn about how, when, and where planet formation and migration occurs.

\subsection{Current State of Obliquity Measurements}

We can measure the orbital obliquities of exoplanets using a variety of methods, but the most successful to date has been time-series high-resolution spectroscopic observations through the transit, and analysis of the Rossiter-McLaughlin effect anomaly through radial velocity \citep[e.g.,][]{Winn05} or Doppler tomographic methodology \citep[e.g.,][]{CollierCameron10W33}. Other methods to measure the obliquity, which are generally applicable only to a subset of the population, include the effects of starspots \citep[e.g.,][]{Dai16} or gravity darkening \citep[e.g.,][]{Barnes09} upon the transit light curves. Alternately, the obliquity can be inferred for some transiting systems by measuring the inclination of the stellar rotation axis with respect to the line of sight using asteroseismology \citep[e.g.,][]{Chaplin13}, rotational variability \citep[e.g.,][]{Mazeh15,LiWinn16}, or the stellar rotation velocity \citep[e.g.,][]{Schlaufman10}. These methods are typically less observationally expensive than transit spectroscopy, but are often subject to caveats.

Observations to date have measured the orbital obliquities of over one hundred planets, 
the vast majority of which are hot Jupiters that have been observed spectroscopically; only a handful of smaller planets have published obliquity measurements \citep{Hirano11,SanchisOjedaWinn11,Albrecht13,Bourrier18,Zhou18}. The available measurements for hot Jupiters have allowed us to identify a number of patterns in their obliquity distributions and make inferences about their dynamical histories and the effects of tides upon these planets \citep[e.g.,][]{Winn10,Hebrard11,MortonJohnson11,Dawson14}. For example, hot Jupiters around A and early F stars have a much wider range of obliquities than those around late F, G, and K stars, which is thought to be do to higher efficiency of tidal damping for the latter systems \citep{Winn10,Dawson14}. The sample for small planets is insufficient to pick out any trends, except that compact multiplanet systems have less oblique orbits than hot Jupiters typically do \citep{Albrecht13}. {\bf Measuring the obliquities of a large sample of small planets will allow us to trace the dynamical histories of the most common types of planetary systems.}

\section{Motivations}

\textbf{Motivation 1} \emph{A window into the dynamical histories of planetary systems}  

Measuring the obliquities of individual planets within multi-planetary systems can help us probe their overall dynamical histories through characterizing their mutual inclinations. The mutual inclinations of planetary systems encode both planet-disk and planet-planet interactions. For compact super-Earth systems, the mutual inclination distribution can yield insights into the gas environments in which they were formed \citep{Dawson15b}. Planet-planet interactions post gas-dissipation are also thought to excite the inclinations of close-in planets \citep[e.g.][]{Hansen13}, potentially leading to the large mutual inclinations of systems with close-in companions \citep{Dai18}. Conversely, mutual inclinations and obliquities of resonant pairs of planets could be damped out by coupling between the orbital and spin precession of the planets \citep{MillhollandLaughlin19}.

The Solar System hosts both Jovian and terrestrial planets, and the interplay between the gas giants and the rocky planets contributed to a water rich habitable environment in the inner Solar System \citep{Raymond06b,Raymond06a}. Understanding the interactions between gas giants and small planets in exoplanet systems will be a key step towards finding true Solar System analogues in the 2020s. For compact systems of super-Earths and Neptunes, companion Jovian planets are thought to excite the mutual inclination and dynamically destabilize the inner planets \citep{Becker17,Huang17}, while in some scenarios an external perturber could tilt an entire inner planetary system \citep{Kaib11}. In the era of \emph{Gaia}, we expect an abundance of exterior Jovian planets with interior transiting companions that will be suitable for obliquity measurements.

\textbf{We recommend an effort to measure the individual obliquities of planets in multi-planet systems, to build a distribution of mutual inclinations for a variety of planetary systems. }

\vspace{5mm}

\textbf{Motivation 2} \emph{Addressing the origins of super-Earths and Neptunes in close-in orbits}

There are two particular formation pathways that are often thought to produce close-in Neptunes and super-Earths. In the classical scenario, Neptune-like planets are formed beyond the ice line. 
These planets are water-rich -- they were formed far away from the star in a region where water can condense into solid form. 
These planets could then have migrated close to their stars either through tidal interactions between the planets and the protoplanetary disk, or through more dynamically violent interactions with other planets.
Planets that migrated through the disk should have well-aligned orbits, while those
with dynamically chaotic pasts could be found on orbital planes strongly misaligned to their parent star \citep[e.g.][]{Bourrier18}.

In a competing scenario, these super-Earths and Neptunes formed `in-situ' within the inner planetary system close to their host stars, a region where our Solar System is devoid of materials. Around a young star enveloped by a thick protoplanetary disk, there is actually enough gas and dust to build our Solar System 
five
times over \citep{ChiangLaughlin13}. 
Quickly, these protoplanet cores accrete an early thick hydrogen-helium atmosphere, just as the intense young stellar irradiation and magnetic interactions begin to clear a gap in the inner disk \citep[e.g.,][]{ChiangLaughlin13}. We expect this population to be found in predominantly multi-planet systems with low stellar obliquities \citep{Hansen13}. If disk tilting mechanisms are active in a large fraction of planetary systems \citep[e.g.][]{Batygin12}, we also expect a sub-population of multi-planet Neptunian and super-Earths to exhibit low mutual inclinations but high stellar obliquities.

\textbf{We recommend a survey for the obliquities of super-Earths and Neptunes with a variety of orbital periods, about a diverse set of parent stars, to test the `in-situ' and migration formation scenarios.}

\section{Obliquities in the 2020s}

With the advent of the wide field space-based transiting planet searches \emph{TESS} and \emph{PLATO}, and continued results from archival data from \emph{Kepler} and \emph{K2}, the coming years will see a revolution in the number of known small transiting planets, particularly those around bright stars. \tess\ will provide nearby planets with orbital periods of days to weeks, while \emph{PLATO} will find nearby planets with periods up to hundreds of days. Planets transiting bright stars are amenable to detailed characterization, including the measurement of their orbital obliquities. The higher signal-to-noise attainable for these stars will allow current facilities to push down to Neptune-sized planets \citep[e.g.,][]{Zhou18}.

Over the coming decade, the next generation of ground-based facilities will come online. The Giant Magellan Telescope (GMT) and the Thirty Meter Telescope (TMT) will have unprecedented light gathering power that are coupled with state-of-art highly stabilized high resolution spectroscopic instrumentation \citep[e.g.][]{Szentgyorgyi14}. These facilities offer new possibilities to explore the obliquities of small transiting planets that are out of reach of today's instrumentation. 

\begin{figure}[h]
  \centering
  \begin{tabular}{ccc}
    $5.0\,R_\oplus$ with the 10\,m Keck & $4.5\,R_\oplus$ with GMT & $1.6\,R_\oplus$ with GMT  \\
    \includegraphics[width=4.5cm]{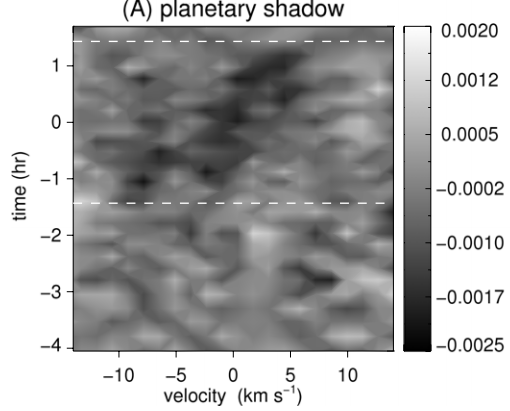} & \includegraphics[width=5cm]{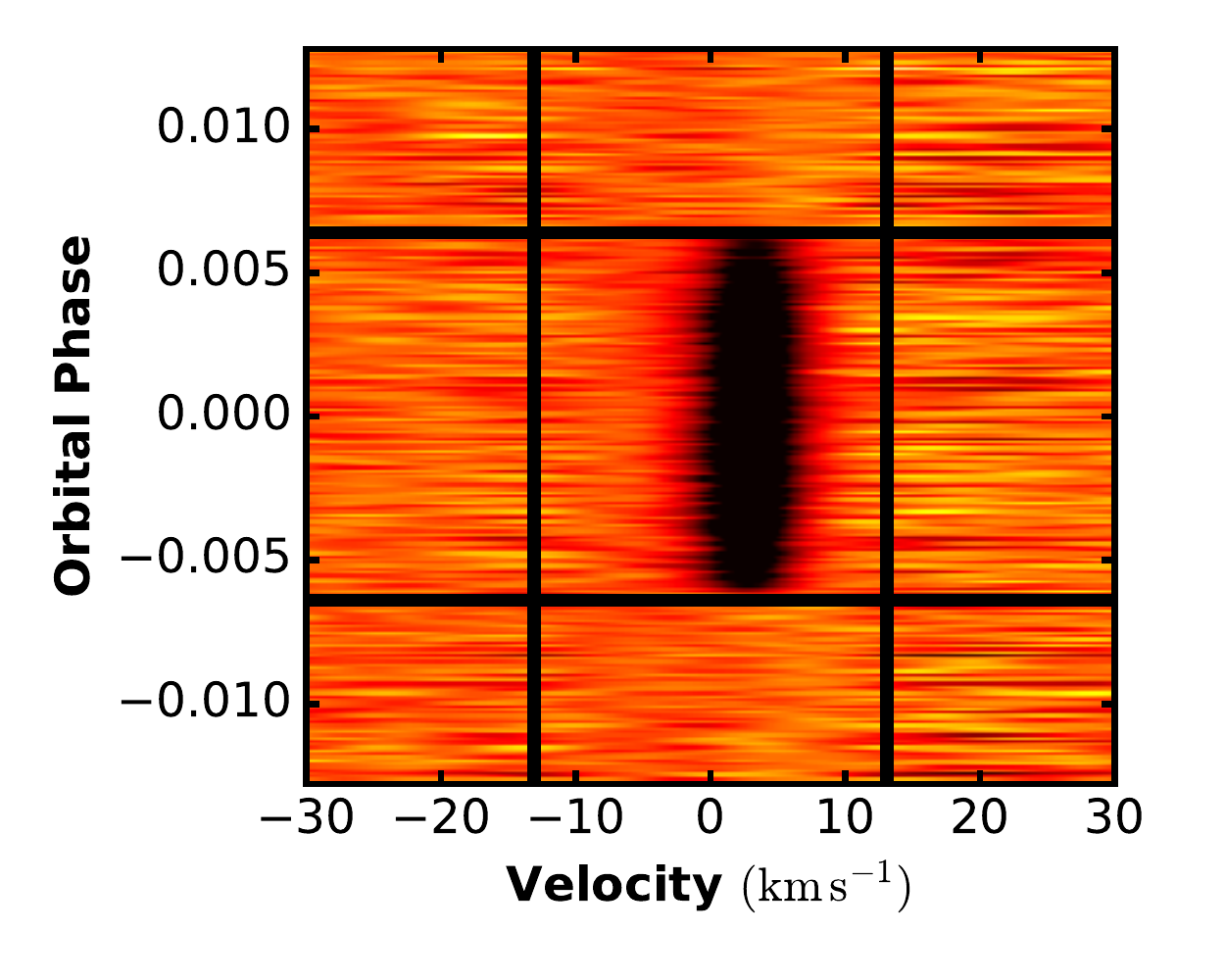} & \includegraphics[width=5cm]{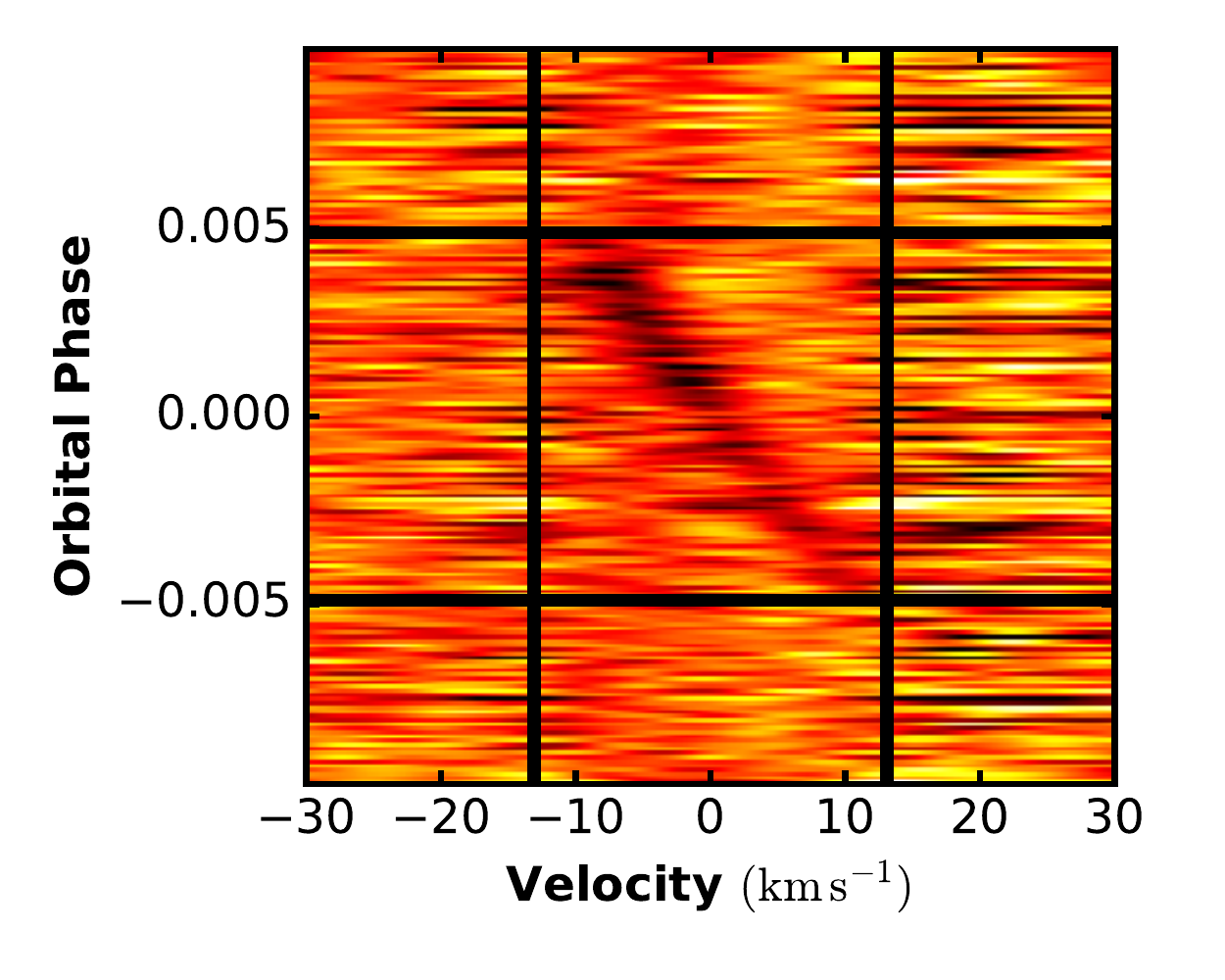} \\
  \end{tabular}
  \caption{Measuring the orbital obliquities of small planets via spectroscopic transits. \textbf{Left} The KECK/HIRES transit spectroscopic detection of the $5\,R_\oplus$ prograde warm Neptune Kepler-25c \citep{Albrecht13}, one of the shallowest spectroscopic transit detections of a planet to date. With $V_\mathrm{mag}=10.8$, Kepler-25 is among the brightest \kepler\ planet hosts, but \tess\ will find many planets around even brighter stars. \textbf{Middle} The transit of a $4\,R_\oplus$ warm Neptune around a $V_\mathrm{mag} = 9$ star, in a polar orbit, simulated for GMT/G-CLEF. \textbf{Right} A simulated transit of a $1.6\,R_\oplus$ super-Earth in a retrograde orbit about a $V_\mathrm{mag} = 9$ star with GMT/G-CLEF. }
  \label{fig:ELT}
  \end{figure}

Figure~\ref{fig:ELT} shows a set of simulated GMT/G-CLEF transit observations to demonstrate the capabilities of the ELTs in measuring the obliquities of small planets via individual transit 
events. The middle and right panels show the transits of a $4\,R_\oplus$ Neptune and a $1.6\,R_\oplus$ super-Earth orbiting a $V_\mathrm{mag} = 9$ F star. 
These simulations are made by scaling existing transit observations from Magellan/MIKE \citep[from][]{Zhou18} to the expected light gathering power of GMT G-CLEF. Similar results should be expected of future high resolution optical echelles on TMT (e.g. HROS), while high-resolution, wide-bandwidth infrared spectrographs on these telescopes (GMT/GMTNIRS, TMT/MODHIS or NIRES) will allow obliquity measurements for small planets around very low-mass or young stars.

\begin{figure}[h]
    \centering
    \vspace{-3mm}
    \centering
    \includegraphics[width=0.45\textwidth]{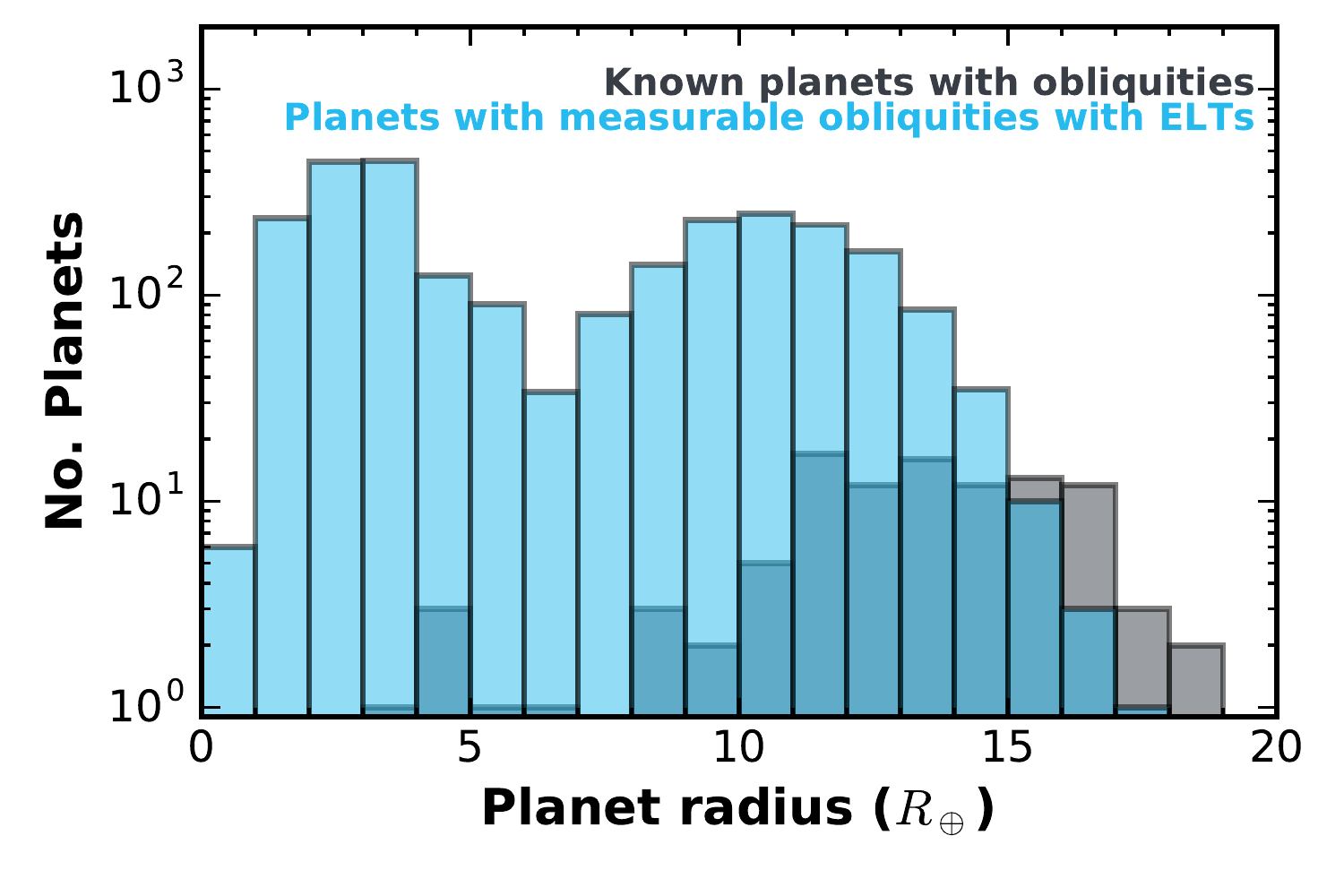}
    \includegraphics[width=0.45\textwidth]{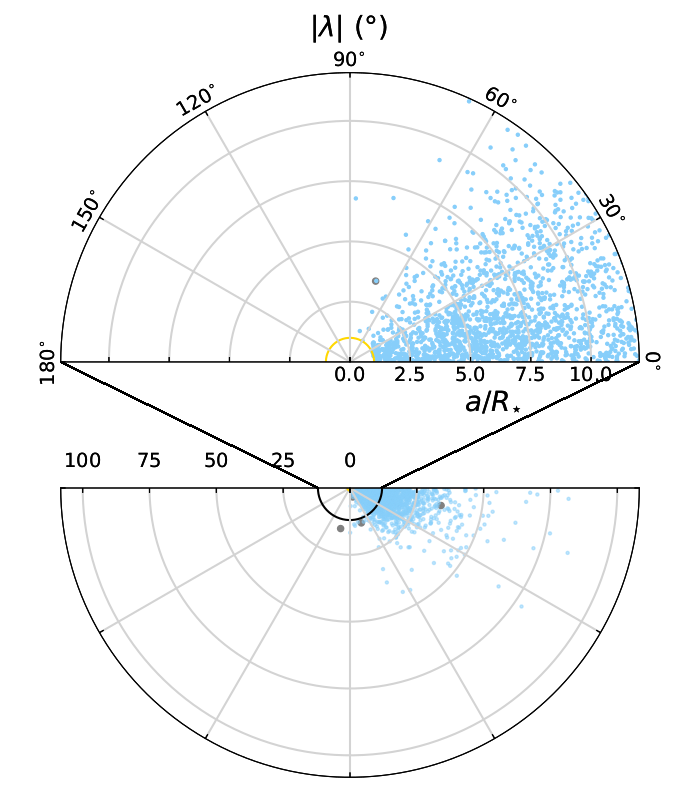}
        \caption{{\bf Left} Histogram showing the number of planets with measured obliquities today (black), against 
        those with obliquities measurable with the ELTs (blue). Few planets smaller than Neptune have orbital obliquities measured by today's facilities. The ELTs will probe the obliquities of planets as small as Earth, opening a new regime for understanding the evolution of planetary systems. {\bf Right} Existing obliquity measurements for super-Earths and Neptunes (gray) along with anticipated \tess\ planets detectable with the ELTs (blue); the obliquities of these planets are drawn from a Fischer distribution \citep[see][]{FabryckyWinn09} with a dispersion of $20^{\circ}$. Each panel shows the sky-projected obliquity $|\lambda|$ as a function of the scaled semi-major axis $a/R_{\star}$. The top panel shows close-in planets ($a/R_{\star}<12$), while the bottom panel mirrors this with a zoomed-out view of the full sample. A gold circle depicts the stellar surface at $a/R_{\star}=1$, while the black circle in the lower panel corresponds the outer boundary of the top panel.  
    \label{fig:tess}}
    \vspace{-3mm}
\end{figure}

Hundreds of small planets will be suitable for spectroscopic obliquity measurements in the near future. Missions like \emph{TESS} and \emph{PLATO} are designed to deliver small planets around bright stars that are suitable for follow-up observations. 
Adopting the simulation above, we estimate that 40 super-Earths ($R_p < 1.6\,R_\oplus$) and 440 Neptunes ($1.6 < R_p < 4 \, R_\oplus$) will be discovered during the primary \emph{TESS} mission that will be suitable for obliquity measurements with the ELTs. Figure~\ref{fig:tess} shows the distributions of these planets in comparison to those with obliquity measurements today. We fully expect this to be a conservative estimate in the anticipation of \emph{TESS} extended missions and the wealth of planets from \emph{PLATO} over the next decade.

A sample of hundreds of obliquity measurements is sufficient to search for correlations between the obliquity distributions and the properties of the system that could affect the formation and dynamics, such as stellar mass, age, multiplicity, or the presence of giant planets or additional stars in the system.

\section{Synergies with Other High-Resolution Spectroscopy Science Cases}

A variety of other observations of these exoplanetary systems are necessary to perform a full dynamical characterization of these systems, and to fully interpret the obliquity measurements.
Obliquity observations are highly complimentary to measurements of the planetary masses (see companion White Paper on radial velocity measurements by D.~R.~Ciardi et al.) and atmospheric composition (see companion White Paper by D.~Dragomir et al.), and observations of the same systems will enable better characterization of the overall system dynamics than either can alone. Does the obliquity distribution differ for planets of different densities and compositions? This would indicate that the same processes (e.g., giant impacts) shape both obliquities and densities. Furthermore, the radial velocity detection of additional non-transiting planets in a system could affect the interpretation. Do systems with non-transiting planets (i.e., larger mutual inclinations, also affected by dynamics) exhibit a different obliquity distribution than nearly-coplanar multi-transiting systems? Finally, combining obliquity information with information on the atmospheric compositions can distinguish planets that formed in situ, or migrated in via disk interactions versus dynamically chaotic migration scenarios.

\section{Conclusions/Recommendations}

We endorse the findings and recommendations published in the National
Academy reports on Exoplanet Science Strategy and Astrobiology Strategy for
the Search for Life in the Universe. This white paper extends and complements
the material presented therein. 
A program to measure the orbital obliquities of a
 diverse sample of hundreds of super-Earths and Neptunes 
will provide unprecedented information on planet formation, migration, and evolution. By dynamically characterizing the most common types of exoplanets and exoplanetary systems, we will be able to study the dominant modes of planet formation and migration. This program will take advantage of the large number of transiting planets around bright stars that will be found by the \tess\ and \emph{PLATO} missions. Such a program will require a large investment of time on both existing facilities and the ELTs; the latter will be essential in order to measure the obliquities of a large number of super-Earths. Overall this is an ambitious but achievable program that will offer unique contributions towards understanding the formation and evolution of planetary systems.

\pagebreak

\bibliography{bibmaster}{}
\bibliographystyle{apj}

\end{document}